# Randomness Testing of Compressed Data

Weiling Chang[1]*, Binxing Fang[1,2], Xiaochun Yun[2], Shupeng Wang[2], Xiangzhan Yu[1]

**Abstract**—Random Number Generators play a critical role in a number of important applications. In practice, statistical testing is employed to gather evidence that a generator indeed produces numbers that appear to be random. In this paper, we reports on the studies that were conducted on the compressed data using 8 compression algorithms or compressors. The test results suggest that the output of compression algorithms or compressors has bad randomness, the compression algorithms or compressors are not suitable as random number generator. We also found that, for the same compression algorithm, there exists positive correlation relationship between compression ratio and randomness, increasing the compression ratio increases randomness of compressed data. As time permits, additional randomness testing efforts will be conducted.

**Index Terms**— Compression technologies, Data compaction and compression, Random number generation

——————————  ◆  ——————————

## 1 INTRODUCTION

Random number generators are an important primitive widely used in simulation and cryptography.
Generating good random numbers is therefore of critical importance for scientific software environments.

There are many different ways to test for randomness, but all of them, in essence, boil down to computing a mathematical metric from the data stream being tested and comparing the result with the expectation value for an infinite sequence of genuinely random data. Output from well-designed pseudo-random number generators should pass assorted statistical tests probing for non-randomness.

This paper describes how the output for each of the lossless data compressors was collected and then evaluated for randomness. It discusses what was learned utilizing the NIST statistical and the Diehard tests and offers an interpretation of the empirical results. In Section 2, the randomness testing experimental setup is defined and described. In Section 3, the empirical results compiled to date are discussed and the interpretation of the test results is presented. Lastly, a summary of lessons learned is presented.

## 2 METHODOLOGY

### 2.1 Test files

We carried out random tests on the contents of five corpora: the Calgary Corpus [1], the Canterbury Corpus [1, 2], the Maximum Compression Corpus, the 100MB file enwik8, and the HitIct corpus.

Two well-known data sets, the Calgary corpus and the Canterbury Corpus, are used by researchers in the universal lossless data compression field. Over the years of using of these two corpora some observations have proven their important disadvantages. The most important in our opinion are: the lack of large files and an over-representation of English-language texts. In order to avoid the two disadvantages, we introduced three other data sets: the Maximum Compression Corpus, the 100MB file enwik8, and the HitIct corpus.

The Maximum Compression benchmark [3] is a website maintained by Werner Bergmans. It uses two data sets, one public and one private. The Maximum Compression Corpus is the public data set of MaximumCompression, which consists of about 55 MiB in 10 files with a variety of types: text in various formats, executable data, and images. The enwik8 [4] is the first 100,000,000 characters of a specific version of English Wikipedia. The HitIct corpus [5] is a Chinese Corpus which consists of 10 files derived from the application of Chinese.

Four basic compression algorithms and four popular compressors were tested, namely Huffman coding, arithmetic coding, LZSS and LZW which adapted from the related codes of the data compression book [6], PPMVC [7], WinZip 12.1, WinRAR 3.90 and WinRK 3.12.

We harnessed and analyzed the five different sets of data (compressed using different algorithms/compressors) for each of these algorithms.

### 2.2 Randomness tests

Randomness is a probabilistic property; the properties of a random sequence are characterized and described in terms of probability. There are an infinite number of possible statistical tests, each assessing the presence or absence of a pattern which, if detected, would indicate that the sequence is non-random. Because there are so many tests for judging whether a sequence is random or not, no specific finite set of tests is deemed complete. In this pa-

————————————————
1. Research Centre of Computer Network and Information Security Technology, Harbin Institute of Technology, Harbin 150001, China
2. Institute of Computing Technology, Chinese Academy of Science, Beijing 100080, China
* Correspondence Auther.

Supported by the National High-Tech Development 863 Program of China (Grant Nos.2009AA01A403, 2007AA01Z406, 2007AA010501, 2009AA01Z437)

per, we will focus on the NIST 800-22 statistical test suite and the Diehard test suite.

**The NIST Statistical Test-Suite**

NIST has developed a suite of 15 tests to test the randomness of binary sequences produced by either hardware or software based cryptographic random or pseudorandom number generators. The tests have been documented in NIST Special Publication (SP) 800-22, " A Statistical Test Suite for Random and Pseudorandom Number Generators for Cryptographic Applications" [8]. These tests focus on a variety of different types of non-randomness that could exist in a sequence. The publication and the associated tests are intended for individuals who are responsible for the testing and evaluation of random and pseudorandom number generators, including (P)RNG developers and testers. SP 800-22 provides a high-level description and examples for each of the 15 tests, along with the mathematical background for each test. The 15 tests are listed in Table 1 [14].

TABLE 1. GENERAL CHARACTERISTICS OF NIST STATISTICAL TESTS

| Test Name | Characteristics |
| --- | --- |
| Frequency (Monobit) Test | Too many zeroes or ones. |
| Frequency Test within a Block | |
| Cumulative Sums Test | Too many zeroes or ones at the beginning of the sequence. |
| Runs Test | Large (small) total number of runs indicates that the oscillation in the bit stream is too fast (too slow). |
| Tests for the Longest-Run-of-Ones in a Block | Deviation of the distribution of long runs of ones. |
| Binary Matrix Rank Test | Deviation of the rank distribution from a corresponding random sequence, due to periodicity. |
| Discrete Fourier Transform (Spectral) Test | Periodic features in the bit stream. |
| Non-overlapping Template Matching Test | Too many occurrences of non-periodic templates. |
| Overlapping Template Matching Test | Too many occurrences of m-bit runs of ones. |
| Maurer's "Universal Statistical" Test | Compressibility (regularity). |
| Approximate Entropy Test | Non-uniform distribution of m-length words. Small values of ApEn(m) imply strong regularity. |
| Random Excursions Test | Deviation from the distribution of the number of visits of a random walk to a certain state. |
| Random Excursions Variant Test | Deviation from the distribution of the total number of visits (across many random walks) to a certain state. |
| Serial Test | Non-uniform distribution of m-length words. Similar to Approximate Entropy. |
| Linear Complexity Test | Deviation from the distribution of the linear complexity for finite length (sub)strings. |

The NIST framework is based on hypothesis testing. A hypothesis test is a procedure for determining if an assertion about a characteristic of a population is reasonable. In this case, the test involves determining whether or not a specific sequence of zeroes and ones is random.

For each statistical test, a set of P-values is produced. The P-value is the probability of obtaining a test statistic as large or larger than the one observed if the sequence is random. Hence, small values are interpreted as evidence that a sequence is unlikely to be random. The decision rule in this case states that "for a fixed significance value α, a sequence fails the statistical test if it's P-value < α." A sequence passes a statistical test whenever the P-value ≥ α and fails otherwise. If the significance level α of a test of $H_0$ (which is that a given binary sequence was produced by a random bit generator.) is too high, then the test may reject sequences that were, in fact, produced by a random bit generator (Type I error). On the other hand, if the significance level α of a test of $H_0$ is too low, then there is the danger that the test may accept sequences even though they were not produced by a random bit generator (Type II error). It is, therefore, important that the test be carefully designed to have a significance level that appropriate for the purpose at hand. However, the calculation of the Type II error is more difficult than the calculation of α because many possible types of non-randomness may exist. Therefore, NIST statistical test suite adopts two further analyses in order to minimize the probability of accepting a sequence being produced by a good generator when the generator was actually bad. First, For each test, a set of sequences from output is subjected to the test, and the proportion of sequences whose corresponding P-value satisfies P-value ≥ α is calculated. If the proportion of success-sequences falls outside of following acceptable interval (confidence interval), there is evidence that the data is non-random.

$$p' = \bar{p} \pm k\sqrt{\frac{\bar{p}(1-\bar{p})}{n}}$$

**(1)**

where $\bar{p}$ = 1 - α, $k$ = 3 is the number of standards deviations, and $n$ is the sample size. If the proportion falls outside of this interval, then there is evidence that the data is non-random.

Second, the distribution of P-values is calculated for each test. If the test sequences are truly random, P-value is expected to appear uniform in [0, 1). NIST recommends to $\chi^2$ test by interval between 0 and 1 is divided into 10 sub-intervals. This is the test of uniformity of P-value. The degree of freedom is 9 in this case. Define $F_i$ as number of occurrence of P-value in $i$ th interval, $s$ is the number of sequences, then $\chi^2$ statistics is given as bellow.

$$\chi^2 = \sum_{i=1}^{10} \frac{(F_i - s/10)^2}{s/10} \qquad (2)$$

The P-value of P-values is calculated such that P-value = **igamc** (9/2, $\chi^2$/2), where **igamc** is the incomplete gamma function. If P-value ≥ 0.0001, i.e., the acceptance region of statistics is $\chi^2 \leq$ 33.72, and then the set of P-values can be considered to be uniformly distributed.

**The Diehard Test-Suite**

The Diehard tests are a battery of statistical tests for measuring the quality of a set of random numbers. They were developed by Professor George Marsaglia of Florida State University over several years and first published in 1995 on a CD-ROM of random numbers. The DIEHARD suite of statistical tests [9] consists of 18 tests. These tests are exquisitely sensitive to subtle departures from randomness, and their results can all be expressed as the probability the results obtained would be observed in a genuinely random sequence. Probability values close to zero or one indicate potential problems, while probabilities in the middle of the range are expected for random sequences.

## 3 EMPIRICAL RESULTS & ANALYSIS

In this section, we show the results of statistical test for the 5 corpora. Randomness can be defined only statistically over a long sequence, it is impossible to comment the randomness of a bit sequence with a single bit sample, so we performed the above two tests many times. We constructed five other test files as follows: encoded all the test files using different compression algorithms or compressors, since many compression techniques add a somewhat predictable preface to their output stream, we skip the 1024 bytes of the beginning of the compressed sequence, and then concatenated them into one file. For each statistical test, further analyses are conducted. Table 2 highlights these categories of data.

TABLE 2. CATEGORIES OF DATA

| File Name | Compressor | Size (Bytes) | Number of Sequences* (NIST test) | Number of pieces (Diehard test) | |
|---|---|---|---|---|---|
| ari | Arithmetic Coding | 125646724 | 958 | 10 | |
| huff | Huffman Coding | 126397455 | 964 | 11 | Constructed by merge all files in five corpora compressed using eight different algorithms / compressors. |
| lzssj | LZSS | 92964564 | 709 | 8 | |
| lzwj | LZW | 125083724 | 954 | 10 | |
| zip | WinZip | 67976783 | 518 | 5 | |
| rar | WinRAR | 58233804 | 444 | 5 | |
| pmv | PPMVC | 51596913 | 393 | 4 | |
| winrk | WinRK | 41190604 | 314 | 3 | |

*Except the random excursion (variant) test

### 3.1 Tests with the NIST Statistical Test Suite

The NIST Statistical Test Suite consists of 15 core statistical tests that, under different parameter inputs, can be viewed as 189 statistical tests. Each P-value corresponds to the application of an individual statistical test on a single binary sequence. Randomness testing was performed using the following strategy:

Input parameters such as the sequence length and significance level were set at $2^{20}$ bits and 0.01, respectively. For each binary sequence and each statistical test, a P-value was reported and a success/failure assessment was made based on whether or not it exceeded or fell below the pre-selected significance level. For each statistical test and each test file, two evaluations were made. First, the proportion of binary sequences in a test file that passed the statistical test was calculated. Second, an additional P-value was calculated, based on a $\chi^2$ test applied to the P-values in the entire sample to ensure uniformity. For both measures described above, an assessment was made. A sample was considered to have passed a statistical test if it satisfied both the proportion and uniformity assessments.

### 3.1.1 Frequency (Monobit) Test

For a truly random sequence, any value in a given random sequence has an equal chance of occurring, and various other patterns in the data should be also distributed equiprobably, i.e. have a uniform distribution. The focus of the Monobit test is the proportion of zeroes and ones for the entire sequence. The purpose of this test is to determine whether the number of ones and zeros in a sequence are approximately the same as would be expected for a truly random sequence.

NIST recommends that the Monobit test should be applied first, since this supplies the most basic evidence for the existence of non-randomness in a sequence, specifically, non-uniformity. All subsequent tests depend on passing this test. If the results of this test support the null hypothesis, then the user may proceed to apply other statistical tests.

For this test, the zeros and ones of the input sequences are converted to values of -1 and +1 and are added together to produce:

$$S_n = X_1 + X_2 + \ldots + X_n$$

where $X_i = 2*\ *i - 1$. Fox example, if $= 1100101101$, then n = 10 and $S_n = 1 + 1 + (-1) + (-1) + 1 + (-1) + 1 + 1 + (-1) + 1 = 2$. Table 3 shows the results of NIST Monobit test for eight test files. All eight files are failed. However, the test files generated by arithmetic algorithm (PPMVC use arithmetic coding to encode the actual selected symbol) have higher success rates (0.7380 for arithmetic coding and 0.7817 for PPMVC respectively).

TABLE 3. RESULTS FOR THE UNIFORMITY OF P-VALUES AND THE PROPORTION OF PASSING SEQUENCES

| File/Compressor | C1 | C2 | C3 | C4 | C5 | C6 | C7 | C8 | C9 | C10 | P-VALUE | PROPORTION | Result |
|---|---|---|---|---|---|---|---|---|---|---|---|---|---|
| ari/Arithmetic Coding | 445 | 114 | 83 | 64 | 55 | 40 | 42 | 34 | 43 | 38 | 0.000000 | 0.7422 | Fail |
| | The minimum pass rate is approximately = 0.980356 for a sample size = 958 binary sequences. | | | | | | | | | | | | |
| huff/Huffman Coding | 964 | 0 | 0 | 0 | 0 | 0 | 0 | 0 | 0 | 0 | 0.000000 | 0.0000 | Fail |
| | The minimum pass rate is approximately = 0.980386 for a sample size = 964 binary sequences. | | | | | | | | | | | | |
| lzssj/LZSS | 709 | 0 | 0 | 0 | 0 | 0 | 0 | 0 | 0 | 0 | 0.000000 | 0.0000 | Fail |
| | The minimum pass rate is approximately = 0.978790 for a sample size = 709 binary sequences. | | | | | | | | | | | | |
| lzwj/LZW | 954 | 0 | 0 | 0 | 0 | 0 | 0 | 0 | 0 | 0 | 0.000000 | 0.0000 | Fail |
| | The minimum pass rate is approximately = 0.980336 for a sample size = 954 binary sequences. | | | | | | | | | | | | |
| zip/WinZip | 501 | 6 | 1 | 5 | 2 | 1 | 1 | 0 | 0 | 1 | 0.000000 | 0.0521 | Fail |
| | The minimum pass rate is approximately = 0.976885 for a sample size = 518 binary sequences. | | | | | | | | | | | | |
| rar/WinRAR | 327 | 21 | 9 | 11 | 16 | 15 | 9 | 12 | 9 | 15 | 0.000000 | 0.3964 | Fail |
| | The minimum pass rate is approximately = 0.975834 for a sample size = 444 binary sequences. | | | | | | | | | | | | |
| pmv/PPMVC | 213 | 44 | 31 | 28 | 24 | 13 | 20 | 8 | 5 | 7 | 0.000000 | 0.7684 | Fail |
| | The minimum pass rate is approximately = 0.974943 for a sample size = 393 binary sequences. | | | | | | | | | | | | |
| winrk/WinRK | 234 | 22 | 15 | 7 | 6 | 5 | 6 | 7 | 7 | 5 | 0.000000 | 0.4618 | Fail |
| | The minimum pass rate is approximately = 0.973155 for a sample size = 314 binary sequences. | | | | | | | | | | | | |

Figure 1 depicts the differences between the proportion of zeroes and ones for all the 46 test files. In each figure, the x-axis is the test file and the y-axis represents the proportion difference ( = 1- (count of ones)/(count of zeros)).

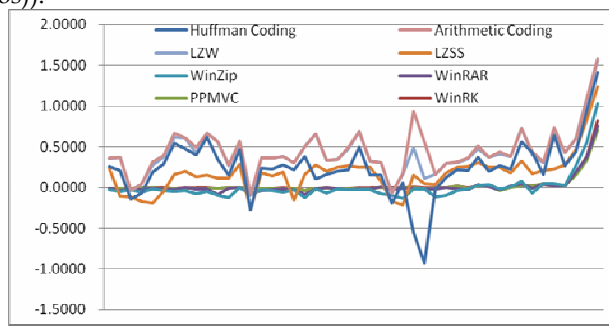

Fig. 1. The differences between the proportion of zeroes and ones (the x-axis indicates the test file, sorted in descending order by file size.)

It can be seen from figure 1, for all the compression algorithms/compressors, the number of ones and zeros in their output are not uniform. We also notice from figure 1 that the output of the WinZip, WinRAR, PPMVC and WinRK is much closer to uniform than other algorithms.

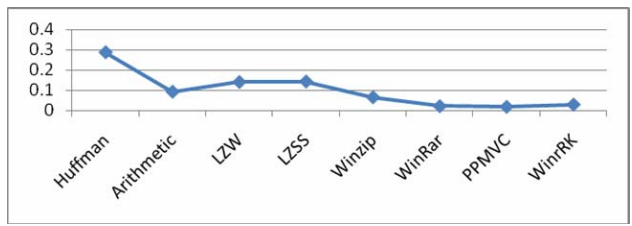

Fig. 2. Standard deviations for proportion differencs (grammar.rk, xargs.rk)

As can be seen from figure 2 that WinZip, WinRAR, PPMVC and WinRK have a lower standard deviation than arithmetic coding, Huffman, LZW and LZSS, it indicates that the their proportion differences tend to be very close to the mean (close to zero), whereas the proportion differences of the arithmetic coding, Huffman, LZW and LZSS are spread out over a large range of values. It seems that the distribution of zeros and ones is much more close to uniform with the increase of the compression ratio.

### 3.1.2 More NIST Test Results

Table 4 shows the results of eight test files. For every test file, the first column shows the P-value's uniformity of each test, the second column shows the passing ratio of each test. All eight test files do not pass the NIST test suite. The two test file pmv and winrk pass 4 tests of all 15 NIST tests, and they seem to be more random than other test files. It is also noted that all 15 tests are not passed for the two test files huff (the output of Huffman Coding) and lzwj (the output of LZW). The reason will be explained later.

TABLE 4. NIST TEST RESULTS OF EIGHT TEST FILES (THE √ SYMBOL DENOTES PASS, 'BLANK' DENOTES FAIL)

| Test Name | ari | | huff | | lzssj | | lzwj | | zip | | rar | | pmv | | winrk | |
|---|---|---|---|---|---|---|---|---|---|---|---|---|---|---|---|---|
| Frequency (Monobit) | | | | | | | | | | | | | | | | |
| BlockFrequency | | | | | | | | | | | | | | √ | | |
| CumulativeSums | | | | | | | | | | | | | | | | |
| Runs | | | | | | | | | | | | | | √ | | |
| LongestRun | | | | | | | | | | | | | | | | |
| Rank | √ | √ | | | √ | √ | | | √ | √ | √ | √ | √ | √ | √ | √ |
| FFT | | | | | | | | | √ | √ | √ | | | | | √ |
| NonOverlappingTemplate | | | | | | | | | | | | | | | | |
| OverlappingTemplate | | | | | | | | | | | | | | | | |
| Universal | | | | | | | | | | | | | √ | √ | √ | √ |
| ApproximateEntropy | | | | | | | | | | | | | | | | |
| RandomExcursions | √ | √ | | | | | | | √ | √ | | | √ | √ | √ | √ |
| RandomExcursionsVariant | √ | √ | | | | | | | √ | | | | √ | √ | √ | √ |
| Serial | | | | | | | | | | | | | | | | |
| LinearComplexity | | | | | √ | √ | | | | | √ | | | | | √ |

Information included in data stream at least has 4 features: *statistical*, *syntax* or *grammar* (the arrangement of symbols to form a message and the structural relationships between these symbols.), *semantics* (which is to do with the range of possible meanings of symbols, dependent on their context the content specify, i.e. its meaning), *pragmatics* (The context wherein the symbols are used. Different contexts can result in different meanings for the same symbols.).

There are two kinds of redundancy contained in the data stream: statistics redundancy and non-statistics redundancy. The non-statistics redundancy includes redundancy derived from syntax, semantics and pragmatics. Order-1 statistics-based compressors compress the statistics redundancy, higher order statistics-based and dictionary-based compression algorithms exploit the statistics redundancy and the non-statistics redundancy. For example, the strong dependency between adjacent symbols of normal text is usually expressed as a Markov model, with the probability of the occurrence of a particular symbol being expressed as a function of the preceding $n$ symbols.

Huffman coding uses a variable-length code table to encode a symbol where the variable-length code table has been derived in a particular way based on the estimated probability of occurrence for each possible value of the symbol. Huffman coding only reduce coding redundancy, it has nothing to do with information redundancy but with the representation of information, i.e., coding itself. Although the binary digits of Huffman coding's output are nearly evenly distributed it still maintains the statistical characteristics of the original data at symbol level, that is, the symbol probability distribution between the variable-length coded symbols of compressed data and the fixed-length coded symbols of original data is identical.

Arithmetic coding is a form of variable-length entropy encoding that converts a string into another representation that represents frequently used characters using fewer bits and infrequently used characters using more bits, with the goal of using fewer bits in total. As opposed to other entropy encoding techniques that separate the input message into its component symbols and replace each symbol with a code word, arithmetic coding encodes the entire message into a single number, a fraction $n$ where (0.0≤ $n$ < 1.0). Arithmetic coders produce near-optimal output for a given set of symbols and probabilities. Compression algorithms that use arithmetic coding (such as PPM, BWT) start by determining a model of the data – basically a prediction of what patterns will be found in the symbols of the message. The model is a prediction algorithm which maintains a statistical model of the data stream. The Huffman coding or the Arithmetic coding is merely a coding scheme, its compression ratio depends on the modeling approach used. The more accurate this prediction is, the closer to optimality the output will be. It can be seen from our experimental results, for Huffman coding and Arithmetic coding, that there is no correlation between their compression ratio and the randomness of their output.

In LZSS, the encoded file consists of a sequence of items, each of which is either a single character (*literal*) or a pointer of the form (*index, length*), the probability distribution of index values is near uniform. LZSS undermines the statistical characteristics contained by the original data stream. However, LZSS can produce new statistical characteristics for the compressed data. The literal values have the characteristics of uneven probability distribution which is different from original data, and so do the length values.

LZW builds a string translation table from the text being compressed. The string translation table maps fixed-length codes to strings. LZW replaces strings of characters with single codes. Under LZW, the compressor never outputs single characters, only phrases. LZW altered the statistical characteristics held by the original data stream. However, by the characteristics of original data, which include data locality and semantic or syntactic attribute, the compressed data produces new statistical characteristics and KCC attribute stemmed from the original data stream.

Altogether, it can be concluded from the experimental results that the output from all the lossless compression algorithms/compressors has bad randomness and increasing the compression ratio can increase the randomness of compressed data.

### 3.2 Tests with the Diehard Test-Suite

The DIEHARD has 18 tests and each test has some P-value. The number of P-Value is different between each test. The sort of test and its P-value is in following table.

| Reference Number | Test Name | Symbol | Number of P-value |
|---|---|---|---|
| 1 | Birthday Spacings Test | BDAY | 10 |
| 2 | Overlapping 5-Permutation Test | OPERM5 | 2 |
| 3 | Binary Rank Test for 31x31 Matrices | RANK31x31 | 1 |
| 4 | Binary Rank Test for 32x32 Matrices | RANK32x32 | 1 |
| 5 | Binary Rank Test for 6x8 Matrices | RANK6x8 | 26 |
| 6 | Bitstream Test | BITSTREAM | 20 |
| 7 | Overlapping-Pairs-Sparse-Occupancy | OPSO | 23 |
| 8 | Overlapping-Quadruples-Sparse-Occupancy | OQSO | 28 |
| 9 | DNA Test | DNA | 31 |
| 10 | Count-The-1's Test on a Stream of Bytes | C1STREAM | 2 |
| 11 | Count-The-1's Test for Specific Bytes | C1BYTE | 25 |
| 12 | Parking Lot Test | PARKLOT | 11 |
| 13 | Minimum Distance Test | MINDIST | 1 |
| 14 | 3D-Spheres Test | 3D | 21 |
| 15 | Squeeze Test | SQEEZE | 1 |
| 16 | Overlapping Sums Test | OSUM | 11 |
| 17 | Runs Test | RUNS | 4 |
| 18 | Craps Test | CRAPS | 2 |

The Diehard test suite was run on a file of at least 80 million bits, so we split our test files into pieces of 11,468,800 bytes. The column of number of pieces in table 2 highlights the split results. There are 220 P-value in a set of DIEHARD so that total number of P-Value is 56 × 220 = 12320 because we test 56 times (The 8 test files split into 56 pieces).

Although the Diehard test suite is one of the most comprehensive publically available sets of randomness tests, unfortunately passing the Diehard tests is not very well defined since Dr. Marsaglia does not provide concrete criteria. Intel [10] assumed that a test is considered failed if it produces a P-value less than or equal to 0.0001 or greater than or equal to 0.9999. It results in a 95% confidence interval of P-values between 0.0001 and 0.9999. This method was used for our testing. The Diehard test results are summarized in Table 6. If multiple P-values are in those results, the worst case value is presented.

TABLE 6: DIEHARD TEST RESULT SUMMARY

| Test Name | P-value | | | | | | | |
|---|---|---|---|---|---|---|---|---|
| | huff | ari | lzwj | lzssj | zip | rar | pmv | winrk |
| BDAY | 1.000000 | 1.000000 | 1.000000 | 1.000000 | 1.000000 | 0.006078 | 0.002076 | 1.000000 |
| OPERM5 | 1.000000 | 1.000000 | 1.000000 | 1.000000 | 1.000000 | 0.016173 | 0.986673 | 0.002675 |
| RANK31x31 | 1.000000 | 1.000000 | 1.000000 | 0.896222 | 1.000000 | 0.780829 | 0.859391 | 0.991590 |
| RANK32x32 | 1.000000 | 1.000000 | 1.000000 | 0.940968 | 1.000000 | 0.829265 | 0.697502 | 0.965003 |
| RANK6x8 | 1.000000 | 1.000000 | 1.000000 | 1.000000 | 1.000000 | 0.999608 | 0.998732 | 1.000000 |
| BITSTREAM | 1.00000 | 1.00000 | 1.00000 | 1.00000 | 1.00000 | 1.00000 | 0.00971 | 1.000000 |
| OPSO | 1.0000 | 1.0000 | 1.0000 | 1.0000 | 1.0000 | 1.0000 | 0.9972 | 0.9988 |
| OQSO | 1.0000 | 1.0000 | 1.0000 | 1.0000 | 1.0000 | 1.0000 | 0.0045 | 0.9988 |
| DNA | 1.0000 | 1.0000 | 1.0000 | 1.0000 | 1.0000 | 1.0000 | 0.0009 | 0.9949 |
| C1STREAM | 1.000000 | 1.000000 | 1.000000 | 1.000000 | 1.000000 | 1.000000 | 0.999301 | 1.000000 |
| C1BYTE | 1.000000 | 1.000000 | 1.000000 | 1.000000 | 1.000000 | 1.000000 | 0.001038 | 1.000000 |
| PARKLOT | 1.000000 | 0.000505 | 1.000000 | 1.000000 | 0.985802 | 0.994722 | 0.009936 | 0.007758 |
| MINDIST | 1.000000 | 1.000000 | 1.000000 | 1.000000 | 1.000000 | 1.000000 | 0.924820 | 1.000000 |
| 3D | 1.000000 | 0.00131 | 1.000000 | 1.000000 | 0.00000 | 0.00089 | 0.00809 | 1.000000 |
| SQEEZE | 1.000000 | 1.000000 | 1.000000 | 1.000000 | 1.000000 | 0.959771 | 0.142881 | 0.999814 |
| OSUM | 0.001560 | 0.012183 | 0.001560 | 0.991762 | 0.001516 | 0.994038 | 0.002129 | 0.976676 |
| RUNS | 1.000000 | 0.001407 | 1.000000 | 1.000000 | 0.975837 | 0.977231 | 0.007877 | 0.025155 |
| CRAPS | 1.000000 | 0.998402 | 1.000000 | 1.000000 | 0.999989 | 0.996809 | 0.919484 | 0.976432 |

We can see from table 6 above, the test file huff, lzwj, lzssj, ari and zip only pass 1 to 3 tests of the 18 Diehard tests, i.e., the output of Huffman coding (Winzip is a combination of LZ77 and Huffman coding), Arithmetic Coding, LZSS and LZW is non-random. This is consistent with the NIST test. The test file rar, winrk and pmv pass most of the tests in Diehard. In order to further investigate their randomness, we must consider the distribution of the P-value.

As far as the P-value is concerned, the central limit theorem does not apply, and large samples do not converge in probability. For a good random number generator, P-values from tests will be uniformly distributed. The distribution of the P-values from the Diehard suite of tests is shown below.

| P-value range | Observed Percent | | | "Expected" Percent |
|---|---|---|---|---|
| | rar | pmv | winrk | |
| 0.0 -- 0.1 | 23.09 | 11.02 | 8.03 | 10 |
| 0.1 -- 0.2 | 5.55 | 8.52 | 6.52 | 10 |
| 0.2 -- 0.3 | 5.27 | 7.27 | 5.76 | 10 |
| 0.3 -- 0.4 | 4.45 | 9.09 | 6.97 | 10 |
| 0.4 -- 0.5 | 5.27 | 8.98 | 8.48 | 10 |
| 0.5 -- 0.6 | 5.45 | 11.14 | 8.03 | 10 |
| 0.6 -- 0.7 | 5.55 | 11.59 | 10.76 | 10 |
| 0.7 -- 0.8 | 7.00 | 9.32 | 9.24 | 10 |
| 0.8 -- 0.9 | 10.18 | 11.36 | 11.36 | 10 |
| 0.9 -- 1.0 | 28.18 | 11.7 | 24.85 | 10 |

Uniformity may also be determined via an application of a $\chi^2$ test. Table 8 shows the valuation of P-value's uniformity of the 3 test files, the statistics is given by equation (2).

TABLE 8: CHI-SQUARE TEST RESULTS OF P-VALUES

Acceptance Region: $\chi^2 \leq 33.72$

| Test File | $\chi^2$ | Result |
|---|---|---|
| rar | 711.45 | Fail |
| pmv | 19.00 | Success |
| winrk | 180.09 | Fail |

Each Diehard test produces one or more P-values. A P-value can be considered good, bad, or suspect. To investigate the randomness of different test files some kind of overall quality metric is needed to convert the sets of P-values produced by test batteries into a single measure, allowing relative comparisons. Meysenburg et al [11, 12, 13] proposed a scheme which assigns a score to a P-value as follows: if $p \geq 0.998$ or $p \leq 0.002$ then it is classified as bad, if $0.95 \leq p < 0.998$ or $0.002 \leq p < 0.05$ then it is classified as suspect. All other P-values are classified as good. The formula used to calculate scores is:

$$s = \sum_{i=1}^{n} \begin{cases} 4 & \text{if } p_i = 0 \vee p_i = 1 \\ 2 & \text{if } p_i \geq 0.998 \vee p_i \leq 0.002 \\ 1 & \text{if } 0.95 \leq p_i < 0.998 \vee 0.002 < p_i \leq 0.05 \\ 0 & \text{otherwise} \end{cases}$$

For each test file, the scores for each test were summed, and the total for each test file is the sum of all the test scores for that test file. Using this scheme, high scores indicate a poor randomness and low scores indicate a good randomness. The results for each test are given in table 9. If the test file is split into multiple pieces, the worst piece is presented.

TABLE 9. DIEHARD TEST RESULTS

| Test Name | Max Score | huff | ari | lzwj | lzssj | zip | rar | pmv | winrk |
|---|---|---|---|---|---|---|---|---|---|
| BDAY | 40 | 40 | 40 | 40 | 40 | 20 | 0 | 0 | 40 |
| OPERM5 | 8 | 8 | 5 | 8 | 8 | 5 | 0 | 2 | 0 |
| RANK31x31 | 4 | 4 | 0 | 4 | 0 | 4 | 0 | 0 | 1 |
| RANK32x32 | 4 | 4 | 0 | 4 | 0 | 4 | 0 | 0 | 1 |
| RANK6x8 | 104 | 104 | 3 | 104 | 104 | 104 | 4 | 7 | 68 |
| BITSTREAM | 80 | 80 | 80 | 80 | 80 | 80 | 53 | 0 | 12 |
| OPSO | 92 | 92 | 92 | 92 | 92 | 92 | 76 | 1 | 6 |
| OQSO | 112 | 112 | 112 | 112 | 112 | 112 | 34 | 3 | 6 |
| DNA | 124 | 124 | 124 | 124 | 124 | 124 | 7 | 5 | 1 |
| C1STREAM | 8 | 8 | 8 | 8 | 8 | 8 | 8 | 0 | 6 |
| C1BYTE | 100 | 100 | 100 | 100 | 100 | 100 | 32 | 7 | 100 |
| PARKLOT | 44 | 38 | 7 | 44 | 44 | 2 | 0 | 1 | 2 |
| MINDIST | 4 | 4 | 4 | 4 | 4 | 4 | 1 | 0 | 4 |
| 3D | 84 | 25 | 1 | 37 | 15 | 16 | 3 | 4 | 39 |
| SQEEZE | 4 | 4 | 0 | 4 | 4 | 4 | 0 | 0 | 2 |
| OSUM | 44 | 3 | 0 | 18 | 1 | 3 | 3 | 0 | 1 |
| RUNS | 16 | 0 | 1 | 11 | 16 | 0 | 0 | 1 | 1 |
| CRAPS | 8 | 8 | 0 | 8 | 5 | 3 | 0 | 0 | 0 |
| **Total** | **880** | **758** | **577** | **802** | **757** | **685** | **221** | **31** | **290** |

It is evident that only the output from the compressor PPMVC (the file pmv) appears to be random for all 18 Diehard tests. Unfortunately, just because a test passes Diehard, that doesn't make it perfect. What has been done is to demonstrate some intrinsic nature of the data compression.

- Increasing the compression ratio increases randomness of compressed data
- Arithmetic coding provides better randomness than other basic compression algorithms (PPMVC use arithmetic coding to code symbols).

## 4 CONCLUSION

The data compressed using 8 different compression algorithms/compressors are tested by means of two popular randomness tests. One is the Special Publication 800-22 issued by the National Institute of Standards and Technology (NIST) and the other is the DIEHARD test provided by Dr. Marsaglia. It is impossible to comment the randomness of a bit sequence with a single bit sample, so we performed the above two tests many times to a data set composed of 5 compression corpora.

The NIST test suite collectively spans many well-known properties that any good cryptographic algorithm should satisfy. All tested files do not pass the NIST test suite. For Diehard test suite, only the pmv file passed. The main conclusion from this investigation is that for the lossless compressed data, there is obvious deviation from randomness, i.e., the output of the lossless compression algorithms/compressors has bad randomness. A secondary conclusion is that, for the same compression algorithm, there exists positive correlation relationship between compression ratio and randomness, increasing the compression ratio increases randomness of compressed data.

**Weiling Chang** was born in Shanxi province, China. He graduated with a BA Econ from University of International Business and Economics (UIBE) in 1993 and earned his master's degree in computer science from China Agricultural University (CAU) in 2006. He is currently in PhD program in Computer Science at Harbin Institute of Technology (HIT), Harbin, China. His major research interests include data compression, computer network and information security.

**Binxing Fang,** born in 1960, is a professor and supervisor of Ph.D. candidates, an academician of Chinese Academy of Engineering and the president of Beijing University of Post and Telecommunications. His research interests include computer network and information security.

**Xiaochun Yun,** born in 1971, is a professor and Ph.D. supervisor at Institute of Computing Technology of the Chinese Academy of Sciences. His research interests include computer network and information security.

**Shupeng Wang,** born in 1980, Ph.D. His research interests include computer network and information security.

**Yu Xiangzhan**,Ph.D.,associate professor of Harbin Institute of Technology,Research Fields: Computer Network and Information Security, Large-Scale Distributed Storage Technology.